\documentclass[review,3p]{elsarticle}

\usepackage{ecrc}

\volume{00}

\firstpage{1}

\runauth{}

\usepackage{amssymb}
\usepackage{float}
\usepackage{amsmath}
\usepackage{bm}
\usepackage{subfigure}

\usepackage[figuresright]{rotating}

\begin{document}

\begin{frontmatter}

\title{High-Spatial-Resolution Monitoring of Strong Magnetic Field using Rb vapor Nanometric-Thin Cell}

\author{G. Hakhumyan${}^{1,2}$ , C. Leroy${}^{2}$, Y. Pashayan-Leroy${}^{2}$, D. Sarkisyan${}^{1}$, M. Auzinsh${}^{3}$}

\address{${}^{1}$Institute for Physical Research, NAS of Armenia, Ashtarak, 0203, Armenia}
\address{${}^{2}$Laboratoire Interdisciplinaire Carnot de Bourgogne, UMR  CNRS 5209 - Universit\'{e} de Bourgogne, F - 21078 Dijon Cedex, France}
\address{${}^{3}$Department of Physics, University of Latvia, 19 Rainis Blvd., Riga LV - 1586, Latvia}

\begin{abstract}
We have implemented the so-called $\lambda$-Zeeman technique (LZT) to investigate individual hyperfine
transitions between Zeeman sublevels of the Rb atoms in a strong external magnetic field $B$ in the
range of $2500 - 5000$ G (recently it was established that LZT is very convenient for the range of $10 - 2500$ G).
Atoms are confined in a nanometric thin cell (NTC) with the thickness $L = \lambda$, where $\lambda$ is the resonant
wavelength 794 nm for Rb $D_1$ line. Narrow velocity selective optical pumping (VSOP) resonances in the transmission
spectrum of the NTC are split into several components in a magnetic field with the frequency positions and transition
probabilities depending on the $B$-field. Possible applications are described, such as magnetometers with nanometric
local spatial resolution and tunable atomic frequency references.
\end{abstract}


\end{frontmatter}

\section{Introduction}
\label{}
A number of optical and magneto-optical processes running at interaction of a narrow-band laser radiation with atomic 
vapors are employed in laser technology, metrology, designing of high-sensitivity magnetometers, problems of quantum 
communications, information storage etc \cite{Budker_1,Budker_2}. As known, the energy levels of atoms undergo frequency 
shifts and changes in their transition probabilities in an external magnetic field $B$. The related effects were studied 
for hyperfine (hf) atomic transitions in optical domain for the transmission spectra obtained with an ordinary cm-size cell 
containing Rb and Cs vapor in \cite{Tremblay_3}. However, because of Doppler broadening (hundreds of MHz), it was possible 
to partially separate different hf transitions only for $B > 1500$ G. Note that even for these large $B$ values the lines 
of the ${}^{85}$Rb and the ${}^{87}$Rb are strongly overlapped, and pure isotopes should be used to avoid complicated spectra. 
In order to eliminate the Doppler broadening, the well-known saturation absorption (SA) technique was implemented to study 
the Rb hf transitions \cite{Momeen_4,Skolnik_5}. However, in this case the complexity of the Zeeman spectra in a magnetic 
field arises primarily from the presence of strong crossover resonances, which are also split into many components. 
That is why, the SA technique is applicable only for $B < 100$ G. Another significant disadvantage is the fact that 
the SA is strongly nonlinear and, therefore, the peak amplitudes of the decreased absorption do not correspond to 
probabilities of atomic transitions at frequencies of which these peaks are formed. This additionally complicates 
the processing of the spectra.
The crossover resonances can be eliminated with selective reflection spectroscopy \cite{Papageorgiou_6}, but to 
correctly determine the hf transition position, the spectra must undergo further non-trivial processing.
It was demonstrated in \cite{Sarkisyan_7,Sarkisyan_8} that the use of resonance fluorescence spectra of a NTC filled 
with Rb atomic vapor and having a thickness of $L = 0.5\lambda$ (where $\lambda = 794$ nm is the wavelength of laser 
radiation whose frequency is resonant to the atomic transition of the $D_1$ line of Rb) allows one to separate and 
to study the atomic transitions between the levels of the hyperfine structure of the $D_1$ line of the ${}^{87}$Rb 
atoms in magnetic fields with $B = 10 - 200$ G. The achieved high sub-Doppler spectral resolution is caused by the 
effect of a strong narrowing of the fluorescence spectrum of a NTC with the atomic vapor column thickness $L = \lambda/2$. 
With the proper choice of the laser intensity and NTC temperature, it is possible to achieve an eightfold narrowing of the 
spectrum compared to that in ordinary $1 - 10$ cm-long cells (for which the Doppler width is about 500 MHz). The method 
was called "half-lambda Zeeman technique" (HLZT). In \cite{Papoyan_9,Sarkisyan_10}, the $D_2$ lines of Rb and Cs atoms 
were studied using HLZT in magnetic fields of about 50 G. Recently it has been shown that the use of the HLZT allows 
one to perform detailed quantitative measurements of both frequency characteristics and probabilities of large number 
of atomic transitions between levels of the hyperfine structure of the Rb $D_1$ line in magnetic fields ranging within 10 -- 2500 G \cite{Hakhumyan_11}.
In \cite{Varzhapetyan_12,Sargsyan_13} it is described a technique based on the use of spectrally-narrow (close to
natural  line-width) velocity selective optical pumping (VSOP) resonances peaks appearing at laser intensities
$\sim 10$ mW/{cm}$^2$ exactly at the positions of atomic transitions in the transmission spectrum of the NTC with 
the Rb vapor column thickness of $L = \lambda$ ($\lambda$ being the wavelength of laser radiation resonant 
with Rb $D_1$ or $D_2$ atomic lines, 794 or 780 nm). Each VSOP resonance is split in an external magnetic field 
into several Zeeman components, the number of which depends on the quantum numbers $F$ of the lower and upper levels. 
The amplitudes of these components and their frequency positions depend unambiguously on $B$-field. Hence, with the 
use of LZT it is also possible to study not only the frequency shift of any individual hf Zeeman transition, but 
also the modification in transition probability. The efficiency of LZT has been proved in the region of 1 - 2500 G \cite{Sargsyan_13}.
Among the most interesting results obtained through HLZT and LZT implementation are vanishing of some of atomic transitions 
for specific $B$-field strength, and the appearance of $\Delta F = 2$ transitions, which are forbidden 
for $B = 0$ \cite{Sargsyan_13}. Moreover, the line intensity of a forbidden transition for some value of 
magnetic field can be higher than that for strong atomic transitions when $B=0$.
It is important to note that LZT could have several advantages as compared with the HLZT: in particular, 
the spectral width of VSOP resonances used in LZT is at least 4 times narrower than that used in HLZT, 
resulting in much higher resolution; also the laser power required for LZT is $\sim 10$ times lower than 
that needed for HLZT; and finally, recording of resonant transmission spectra does not require sensitive 
detectors as in the case of fluorescence spectra.
The aim of the work is to demonstrate that LZT can also be successfully used for even higher external 
magnetic fields up to 5000 G. Possible applications of LZT for diagnostics and mapping of large magnetic 
gradients and for making widely tunable compact frequency references are addressed too.

\section{Experiment}
\subsection{Nanometric-thin cell}
\label{Nanometric-thin cell (NTC)}
The first design of the NTC (called extremely thin cell) consisting of windows and a vertical side-arm 
(a metal reservoir), was presented in \cite{Sarkisyan_14}. Later, this design was somewhat modified and 
a typical example of recent version is presented in \cite{Sarkisyan_15}. The photograph of the NTC with 
smoothly variable thickness wedged in the vertical direction is shown in Fig. \ref{fig:NTC}. The wedged 
gap in this case was formed using a platinum spacer strip of 2 $\mu$m thickness. The presented NTC has 
garnet windows of 0.8 mm thickness (the use of thin wafers in some cases is more convenient), and in 
order to increase the wafers thickness at the bottom to fit the side-arm ($\varnothing =$ 2 mm), two 
additional garnets plates are glued to the main wafers. The NTC is filled with a natural mixture of 
the ${}^{85}$Rb ($72.2\%$) and ${}^{87}$Rb ($27.8\%$). The region of $L \approx \lambda \approx 800$ 
nm is indicated. The temperature limit of the NTC operation is 400 $^\circ$C. The NTC operated with a 
specially designed oven with two ports for laser beam transmission. The source temperature of the atoms 
of the NTC was 120 $^\circ$C, corresponding to the vapor density $N = 2\cdot10^{13}$cm$^{-3}$, but the 
windows were maintained at a temperature that was 20 $^\circ$C higher.

\subsection{Experimental setup}
\begin{figure}[h]
\begin{minipage}[b]{0.47\linewidth}
\centering
\includegraphics[width=3.3in, height=2.57in, keepaspectratio=true]{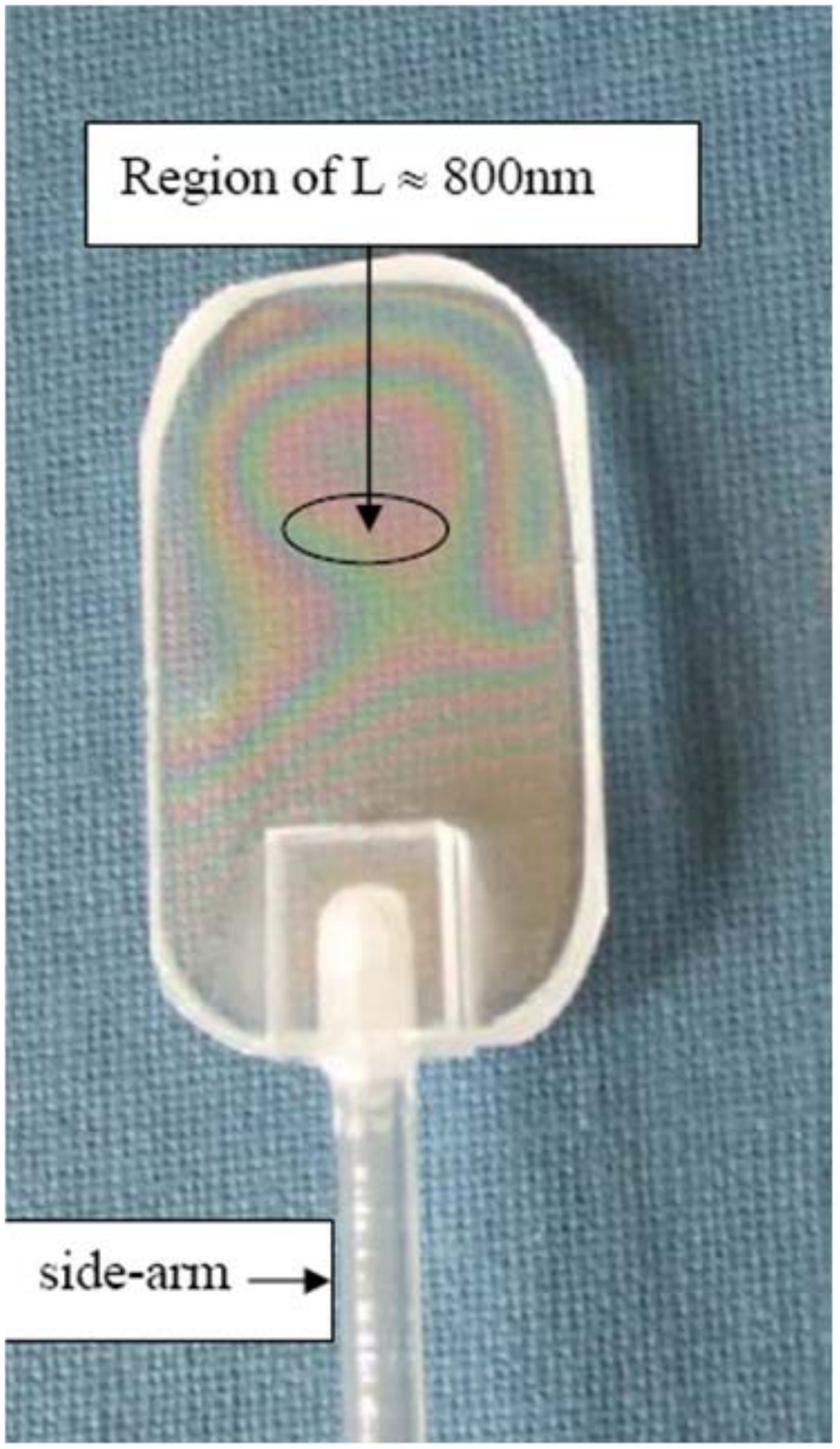}
\caption{Photography of the NTC. Since NTC gap thickness is of order of visible light wavelength, 
interferometric pattern occurs. Region of $L = \lambda = 800$ nm is indicated, NTC has a sapphire side-arm.}
\label{fig:NTC}
\end{minipage}
\hspace{0.6cm}
\begin{minipage}[b]{0.47\linewidth}
\centering
\includegraphics[width=3.3in, height=2.57in, keepaspectratio=true]{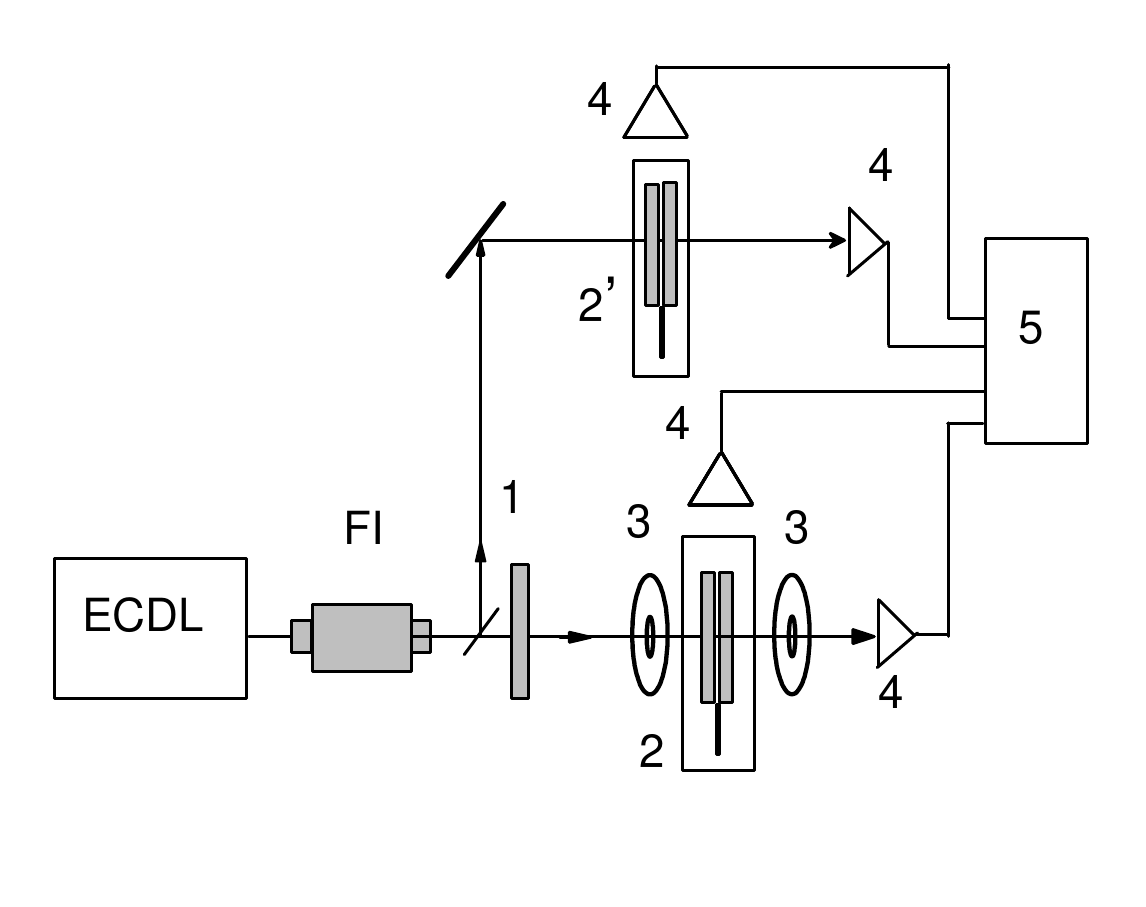}
\caption{Sketch of the experimental setup. ECDL - diode laser, FI - Faraday isolator, $1$ - $\lambda/4$ plate, 
$2$ - NTC in the oven, $2'$ - an auxiliary NTC and the oven, $3$ - ring magnets, $4$ - photodetectors, $5$ - digital storage oscilloscope.}
\label{fig:SetUp}
\end{minipage}
\end{figure}

The sketch of the experimental setup is shown in Fig. \ref{fig:SetUp}. The circularly polarized beam of 
extended cavity diode laser (ECDL, $\lambda = 794$ nm, $P_L = 30$ mW, $\gamma_L < 1$ MHz) resonant with 
the ${}^{87}$Rb $D_1$ transition frequency, was directed onto the Rb NTC (\textit{2}) with the vapor 
column thickness $L = \lambda = 794$ nm at nearly normal incidence. The NTC was inserted in a special 
oven with two openings. The transmission signal was detected by a photodiode (\textit{4}) and was 
recorded by Tektronix TDS 2014B digital four-channel storage oscilloscope (\textit{5}). A Glan prism 
was used to purify initial linear radiation polarization of the laser radiation; to produce a circular 
polarization, a $\lambda/4$ plate (\textit{1}) was utilized. Magnetic field was directed along the 
laser radiation propagation direction $\textbf{k}$ ($\textbf{B}\|\textbf{k}$). About $50 \%$ of the 
pump power was branched by a beam splitter to an auxiliary Rb NTC ($2^{\prime}$).
The fluorescence spectrum of the latter at $L = \lambda/2$ was used as a frequency reference for $B = 0$. 
Moderate longitudinal magnetic field ($B < 250$ G) was applied to the NTC by a system of Helmholtz coils 
(not shown in Fig. \ref{fig:SetUp}). The $B$-field strength was measured by a calibrated Hall gauge. It 
is important to note that the use of the NTC allows one to apply very strong magnetic fields using widely 
available strong permanent ring magnets (PRM): in spite of strong inhomogeneity of magnetic field (in our 
case it can reach 150 G/mm), the variation of $B$-field inside atomic vapor column is $\sim 0.1$ G, i.e. 
by several orders less than the applied $B$ value, thanks to small thickness of the NTC ($L = 794$ nm).

\subsection{Experimental results and discussion}

\begin{figure}[h]
\centering
\subfigure[]{
\resizebox{0.4\textwidth}{!}{\includegraphics{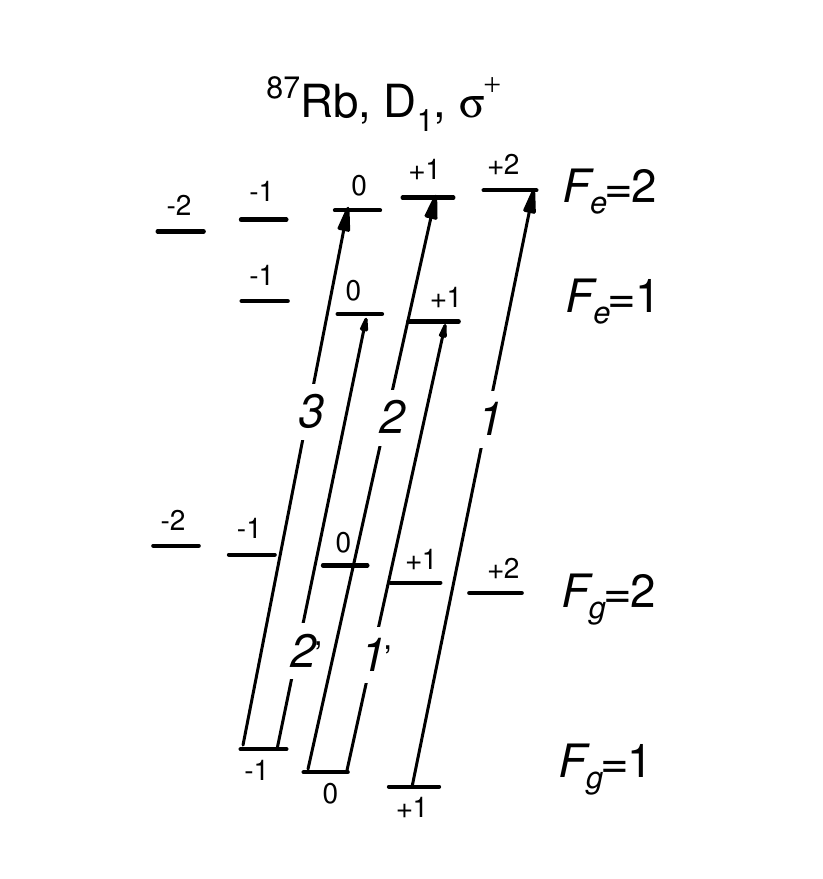}
}
\label{fig:3a}
} \hspace{0.5cm}
\subfigure[]{
\resizebox{0.4\textwidth}{!}{\includegraphics{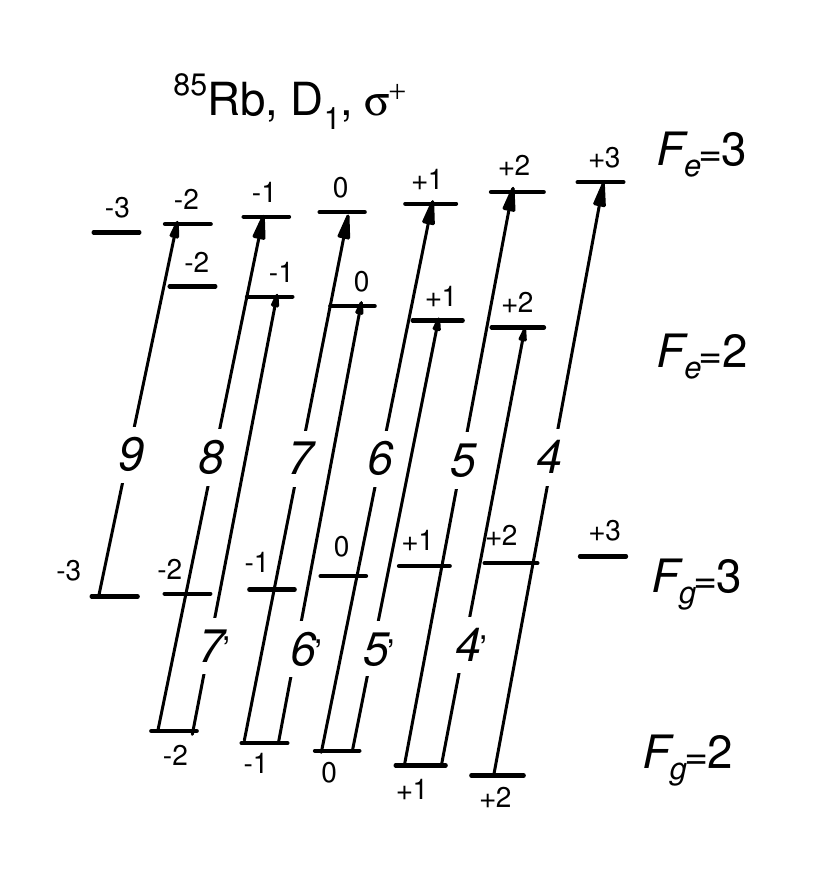}
}
\label{fig:3b}
}
\caption{The hfs energy level diagram of $D_1$ line of ${}^{87}$Rb (a) and ${}^{85}$Rb (b) in magnetic 
field and possible atomic Zeeman transitions for $\sigma^+$ polarized exciting laser radiation.}
\label{fig:fig3}
\end{figure}

\begin{figure}[h]
\includegraphics[width=3.47in, height=2.57in,keepaspectratio=true]{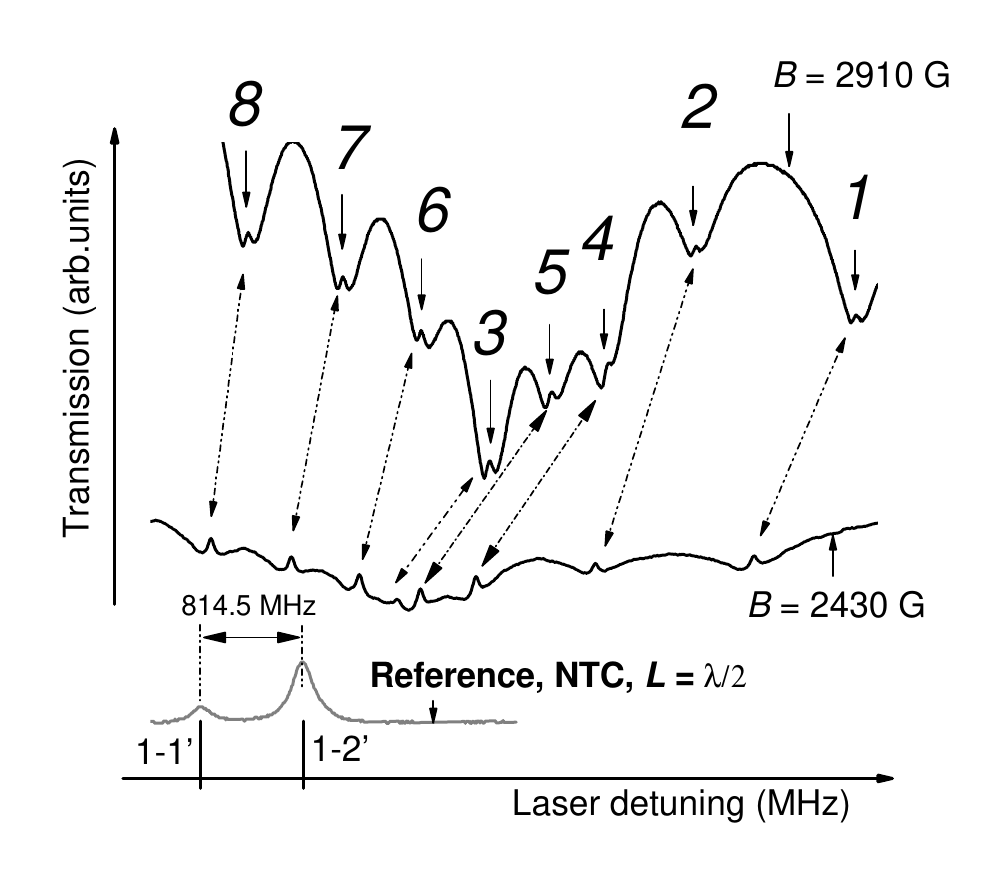}
\centering
\caption{NTC transmission spectra for $L = \lambda$ for the transition between sublevels of hf 
states for the ${}^{85}$Rb and ${}^{87}$Rb $D_1$ line in the case of $\sigma^+$ excitation at $B = 2910$ G 
(the upper curve) and $B = 2430$ G (middle curve). The lower grey curve presents the reference spectrum, 
which is the fluorescence spectrum of the NTC of $L = \lambda/2$, for $B = 0$. Oblique arrows indicate 
the positions of the VSOPs resonance frequencies with the labels $1 - 8$ for $B = 2430$ G and $B = 2910$ G.}
\label{fig:fig4}
\end{figure}

The allowed transitions between magnetic sublevels of hf states for the ${}^{85}$Rb and ${}^{87}$Rb, $D_1$ 
line in the case of $\sigma^+$ (left circular) polarized excitation and selection  rules $\Delta m_F = +1$ 
are depicted in Fig. \ref{fig:fig3} (LZM works well also for $\sigma^-$ excitation).
In \cite{Sargsyan_13} the maximum attainable strength of $B$-field was 2500 G. In order to increase the 
attainable strength of the external magnetic field applied to the atomic vapor contained in NTC, the 
permanent ring magnets embracing the cell should be as close to each other as possible. The main limitation 
for the distance between PRMs is imposed by the longitudinal dimension of the cell oven (see Fig. \ref{fig:SetUp}). 
Earlier, in \cite{Sargsyan_13} the longitudinal dimension of the oven was 4 cm resulting in the maximum 
attainable strength of 2500 G. A new oven was developed specially for high $B$-field applications, with 
the longitudinal dimension of $\approx 2$ cm. This allows us to produce the attainable strength of $B$-field $> 2500$ G.
As it was shown \cite{Sargsyan_13} LZT is based on the use of spectrally-narrow VSOP  resonances appearing 
at laser intensities $\sim 10$ mW/cm$^2$ in the transmission spectrum of the NTC with the thickness 
$L = \lambda$. The VSOP peaks of reduced absorption are exactly at the atomic transitions, and these VSOPs 
are split into several new components, the number of which depends on the quantum numbers $F$ of the lower 
and upper levels, while the amplitudes of the components and their frequency positions depend on $B$-field.
Fig. \ref{fig:fig4} shows the NTC transmission spectra for $L = \lambda$ for the allowed transitions between 
magnetic sublevels of hf states for the ${}^{85}$Rb and ${}^{87}$Rb, $D_1$ line in the case of $\sigma^+$ 
polarized excitation at $B = 2910$ G (the upper curve) and $B = 2430$ G (the middle curve).  The labels $1-8$ 
denote the corresponding transitions between the magnetic sublevels shown in Fig. \ref{fig:fig3}. As shown, 
all the individual Zeeman transitions are clearly detected. The lower grey curve presents the fluorescence 
spectrum of the NTC of $L = \lambda/2$, which is the reference spectrum for the case $B = 0$. 
For convenience the spectra are shifted vertically.
The oblique arrows indicate the positions of the VSOP resonances with the labels $1-8$ for $B = 2430$ and 2910 G. 
As seen from Fig. \ref{fig:fig4} for magnetic field measurement the most convenient is the VSOP peak 
number \textit{1} (${}^{87}$Rb, $F_g = 1$, $m_F = +1 \rightarrow F_e = 2$, $m_F = +2$), since it has the 
largest peak amplitude among transitions \textit{1, 2, 3} of the ${}^{87}$Rb and it is not overlapped 
with any other transition, while having a strong detuning value of $\sim 1.8$ MHz/G versus magnetic field 
strength. It is important to note that LZT allows one to check whether the VSOP resonance (i.e. peak of 
reduced absorption) is a real one or has an artificial/noise nature. For this purpose one should simply 
increase the side-arm temperature by additional 20 -- 30 degrees in order to provide larger absorption 
in the transmission spectrum. The real VSOP must be located exactly at the bottom of the absorption at 
the position of the corresponding atomic transition shifted in the magnetic field, as it is shown on 
the upper curve of Fig. \ref{fig:fig4} (the side-arm temperature is 140 ${}^\circ$C).
As it is mentioned the strong magnetic field is produced by two PRMs ($\varnothing=$ 30 mm), with 
holes ($\varnothing=$ 3 mm) to allow radiation to pass, placed on opposite sides of the nanocell 
oven and separated by a distance varied between  20 mm and 35 mm (see Fig. \ref{fig:SetUp}). 
To control the magnetic field value, one of the magnets was mounted on a micrometric translation 
stage for longitudinal displacement.
\begin{figure}[h]
\includegraphics[width=3.47in, height=2.57in,keepaspectratio=true]{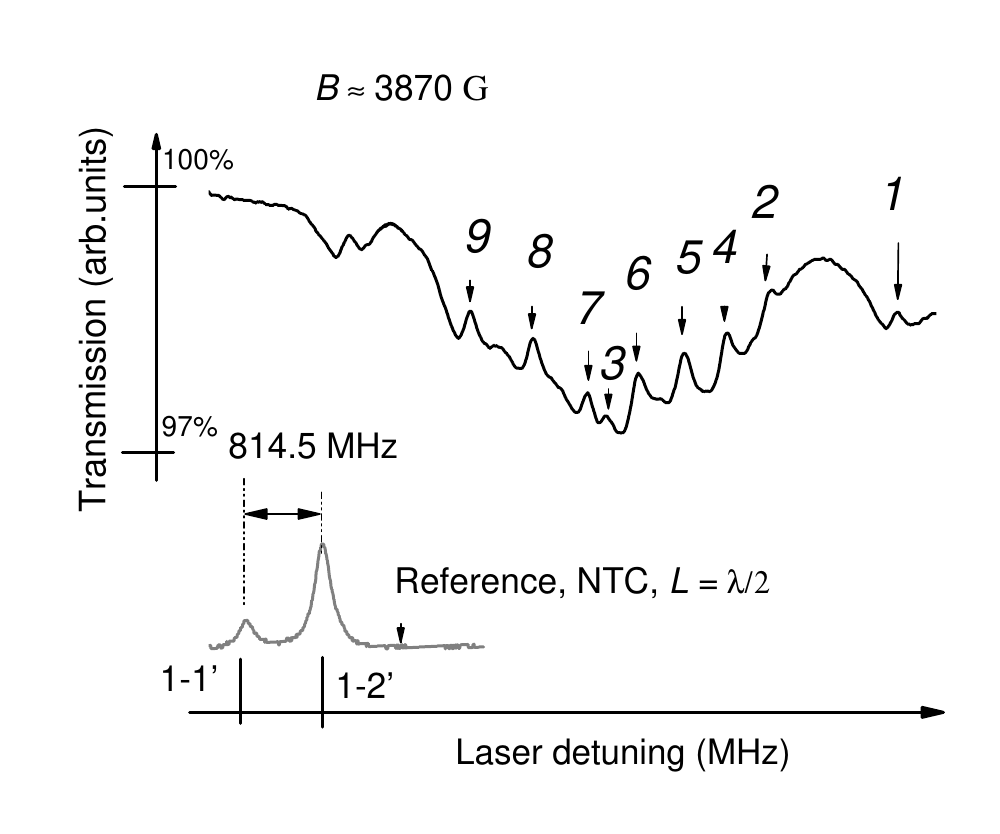}
\centering
\caption{NTC transmission spectra for $L = \lambda$ for the transition between sublevels of hf 
states for the ${}^{85}$Rb and ${}^{87}$Rb $D_1$ line for the case of $\sigma^+$ excitation 
at 3870 G (the upper curve). The lower grey curve presents the reference spectrum, which is 
the fluorescence spectrum of the NTC of $L = \lambda/2$, for $B = 0$.}
\label{fig:fig5}
\end{figure}
The NTC transmission spectra for the thickness $L = \lambda$, for the ${}^{85}$Rb and ${}^{87}$Rb, 
$D_1$ line in the case of $\sigma^+$ excitation at $B = 3870$ G (the upper curve) are presented 
in Fig. \ref{fig:fig5}. The labels $1-9$ denote the corresponding transitions between the magnetic 
sublevels shown in Fig. \ref{fig:fig3}. The lower grey curve presents the fluorescence spectrum of 
the NTC of $L = \lambda/2$, which is the reference spectrum for the case $B = 0$. A new VSOP with 
label \textit{9} is seen in the spectrum (the corresponding atomic transition is shown in 
Fig. \ref{fig:fig3}), while it was absent for the case of smaller $B$-field shown in Fig. \ref{fig:fig4}. 
As it is seen from Fig. \ref{fig:fig5} the most convenient is still the VSOP peak number \textit{1} 
for magnetic field measurement.
The following control experiment was carried out: one of PRMs was set on the table with the 
micrometer step. In the magnetic field $ \sim 4000$ G, one PRM was shifted toward the other by 
the displacement of PRMs by 15 $\mu$m leading to the frequency shift of component \textit{1 } 
by 4 MHz to the high frequency region, which was detected by the comparison of the spectra relatively easy.
\begin{figure}[h]
\includegraphics[width=3.47in, height=2.2in,keepaspectratio=true]{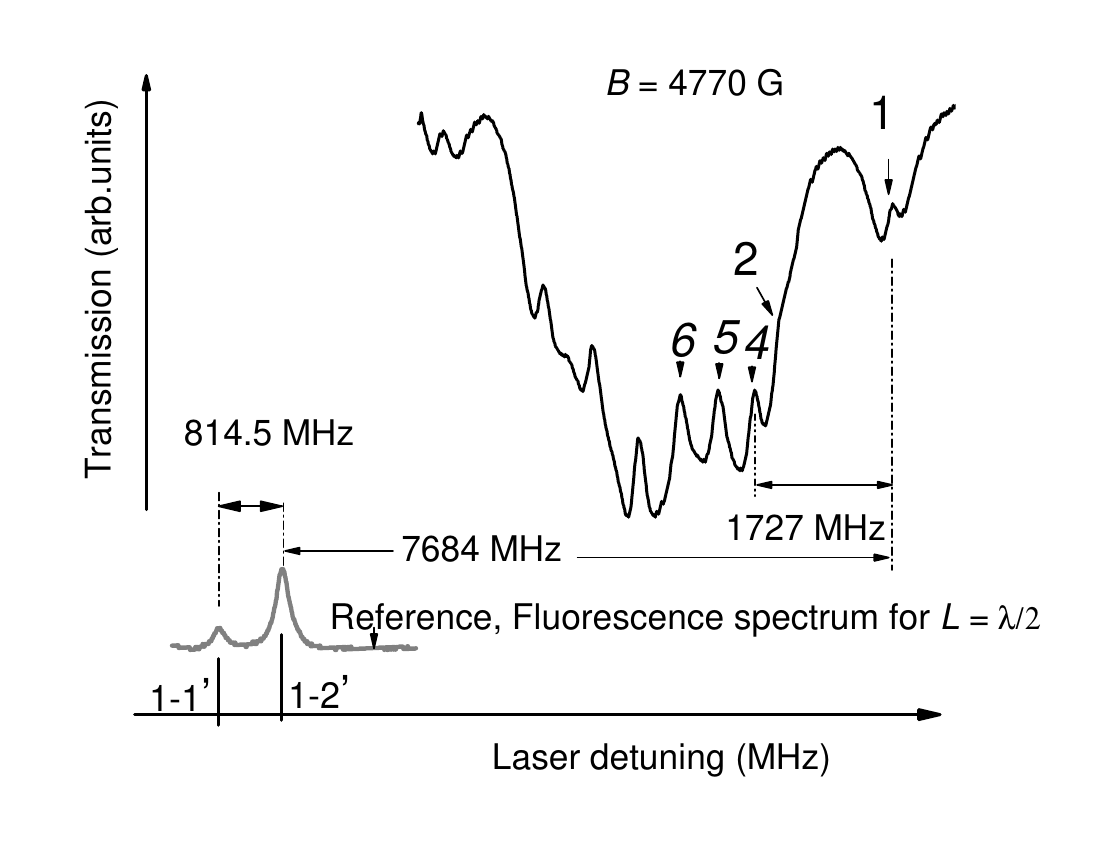}
\centering
\caption{NTC transmission spectra for $L = \lambda$ for the ${}^{85}$Rb and ${}^{87}$Rb $D_1$ line 
in the case of $\sigma^+$ excitation at $B = 4770$ G (the upper curve). The lower grey curve presents 
the reference spectrum, which is the fluorescence spectrum of the NTC of $L = \lambda/2$, for $B = 0$.}
\label{fig:fig6}
\end{figure}
The NTC transmission spectra for thickness $L = \lambda$, for the ${}^{85}$Rb and ${}^{87}$Rb, $D_1$ line 
in the case of $\sigma^+$ excitation at $B = 4770$ G (the upper curve) is presented in Fig. \ref{fig:fig6}. 
The labels $1-6$ denote the corresponding transitions between the magnetic sublevels shown in Fig. \ref{fig:fig3}. 
The lower grey curve presents the fluorescence spectrum of the NTC of $L = \lambda/2$, which is the reference 
spectrum for the case $B = 0$. As it is seen from Fig. \ref{fig:fig6} the VSOP peak number \textit{1} 
is still the most convenient for magnetic field measurement.
We note that transition \textit{1} is strongly shifted by $\sim 7.7$ GHz from the $B = 0$ position of 
the $F_g = 1 \rightarrow F_e = 2$ transition. The latter allows allows of developing a frequency 
reference based on a nanocell and PRMs, widely tunable over a range of several gigahertz onto 
high-frequency wing of transition of ${}^{87}$Rb atom by simple displacement of the magnet.

\section{Theoretical model and discussions}
In this work the main interest is to study the behavior of the  ${}^{87}$Rb, $F_g = 1 \rightarrow F_e = 2$ 
transitions, $D_1$ line in the case of  $\sigma^+$ excitation, since from the application point of 
view they are most interesting. Note, that for $\sigma^-$ excitation the probability of the atomic 
transitions rapidly decreases with magnetic field strength. Also, the probability of the atomic 
transitions ${}^{87}$Rb, $F_g = 1 \rightarrow F_e = 1$ in the case of $\sigma^+$ excitation rapidly 
decreases with magnetic field strength.
Theoretical model describes how to provide the calculations of separated transition's frequencies 
and amplitude modification undergo external magnetic field 
\cite{Tremblay_3,Papageorgiou_6,Sarkisyan_7,Sarkisyan_8,Papoyan_9,Sarkisyan_10,Auzinsh}.
We adopt a matrix representation in the coupled basis, that is, the basis of the unperturbated 
atomic state vectors $\left|(n=3), L, J, F, m_F\right\rangle$ to evaluate the matrix elements 
of the Hamiltonian describing our system. In this basis, the diagonal matrix elements are given by
\begin{equation}
\left\langle F,m_F|H|F,m_F\right\rangle = E_0(F) + \mu_Bg_Fm_FB_Z,
\end{equation}
where $E_0(F)$ is the initial energy of the 
sublevel $\left|(n=3), L, J, F, m_{F}\right\rangle \equiv \left|F, m_F\right\rangle$ 
and $g_F$ is the effective Land\'{e} factor.
The off-diagonal matrix elements are non-zero for levels verifying the selection rules 
$\Delta L = 0, \Delta J = 0, \Delta F = \pm 1, \Delta m_F = 0$,
\begin{equation}
\begin{array}{r}
\left\langle F-1, m_F|H|F, m_F\right\rangle = \left\langle F, m_F|H|F-1, m_F\right\rangle = -\frac{\mu_BB_z}{2}(g_J-g_I)\\
\times \left(\frac{[(J+I+1)^2-F^2][F^2-(J-I)^2]}{F}\right)^{1/2}\left(\frac{F^2-m_F^2}{F(2F+1)(2F-1)}\right)^{1/2}.
\end{array}
\end{equation}
The diagonalization of the Hamiltonian matrix allows one to find the eigenvectors and the eigenvalues, 
that is to determine the eigenvalues corresponding to the energies of  Zeeman sublevels and the new 
states vectors which can be expressed in terms of the initial unperturbed atomic state vectors,
\begin{equation}
\left|\Psi(F_e^{'}, m_{F_e})\right\rangle = \sum_{F_e = I - J_e}^{F_e = I + J_e} \alpha _{F_e^{'}F_e}^e(B_z, m_{F_e})|F_e, m_{F_e}\rangle
\end{equation}
and
\begin{equation}
\left|\Psi(F_g^{'}, m_{F_g})\right\rangle = \sum_{F_g = I - J_g}^{F_g = I + J_g} \alpha _{F_g^{'}F_g}^g(B_z, m_{F_g})|F_g, m_{F_g}\rangle.
\end{equation}
The state vectors $\left|F_e, m_e\right\rangle$ and $\left|F_g, m_g\right\rangle$ are the unperturbated state 
vectors, respectively, for the excited and the ground states.  The coefficients $\alpha _{F_e^{'}F_e}^e(B_z, m_{F_e})$ 
and $\alpha _{F_g^{'}F_g}^g(B_z, m_{F_g})$ are mixing coefficients, respectively, for the excited and the ground states; 
they depend on the field strength and magnetic quantum numbers $m_e$ or $m_g$. Diagonalization of the Hamiltonian 
matrix  for ${}^{87}$Rb, $D_1$ line, in case of $\sigma^+$ polarization of exciting radiation, allows one to 
obtain the shift of position of energy levels in presence of external magnetic field.
\begin{figure}[h]
\includegraphics[width=3.47in, height=2.2in,keepaspectratio=true]{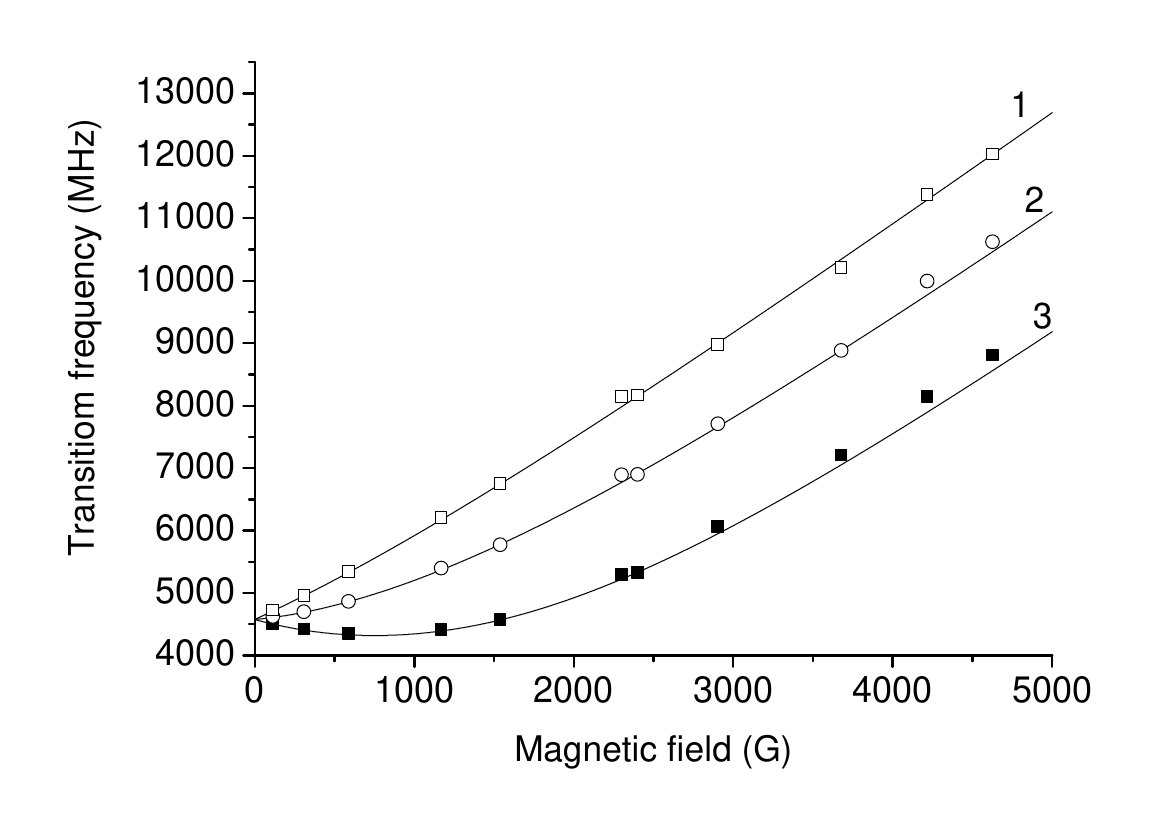}
\centering
\caption{Transition's frequency of components $1$, $2$ and $3$ relatively to the 
initial position at $B = 0$, solid line is the theory.}
\label{fig:fig7}
\end{figure}
Fig. \ref{fig:fig7} shows the frequency shift of components \textit{1, 2} and \textit{3} 
(see diagram in Fig. \ref{fig:3a}) relative to their initial position at $B = 0$.
The probability of a transition, induced by the interaction of the atomic electric dipole 
and the oscillating laser electric field is proportional to the spontaneous emission rate 
of the associated transition $A_{eg}$, that is, to the square of the transfer coefficients 
modified by the presence of the magnetic field
\begin{equation} \label{5}
\frac{8\pi^2}{3\varepsilon_0 \hbar \lambda_{eg}^3}\left|\left\langle e\left|D_q\right|g\right\rangle \right|^2 = 
A_{eg} \propto a^2\left[\Psi (F_e^{'}, m_{F_e}); \Psi(F_g^{'}, m_{F_g});q\right],
\end{equation}
where $D_q$ denote the standard components of the electric dipole moment:
\begin{equation}
\vec{D}.\vec{e} = \sum_q D_q e_q,
\end{equation}
with $q = -1, 0, 1$.
In Eq. \eqref{5}, the transfer coefficients are expressed as
\begin{equation}
a\left[\Psi(F_e^{'}, m_{F_e}); \Psi(F_g^{'}, m_{F_g});q\right] = \sum_{F_e F_g} \alpha_{F_e^{'}F_e}^e 
(B_z, m_{F_e} )a\left(F_e, m_{F_e}; F_g, m_{F_g}; q\right) \alpha_{F_g^{'}F_g}^g (B_z, m_{F_g}),
\end{equation}
where the unperturbated transfer coefficients have the following definition
\begin{equation}
\begin{array}{l} {a\left(F_e, m_{F_e}; F_g, m_{F_g}; q\right) = (-1)^{1 + I + J_e + F_e + F_g - m_{F_e} } } \\
 {\times \sqrt{2J_e + 1} \sqrt{2F_e + 1} \sqrt{2F_g + 1} \left(
 \begin{array}{ccc}
 F_e      & 1 & F_g \\
 -m_{F_e} & q & m_{F_g}
 \end{array}\right)\left\{
 \begin{array}{ccc}
 F_e & 1 & F_g \\
 J_g & I & J_e
 \end{array}\right\}}
\end{array},
\end{equation}
the parenthesis and curly brackets denote, respectively, the $3j$ and $6j$ symbols, $g$ 
and $e$ point respectively ground and excited states.
\begin{figure}[h]
\label{fig:fig8} \centering
\subfigure[]{
\resizebox{0.4\textwidth}{!}{\includegraphics{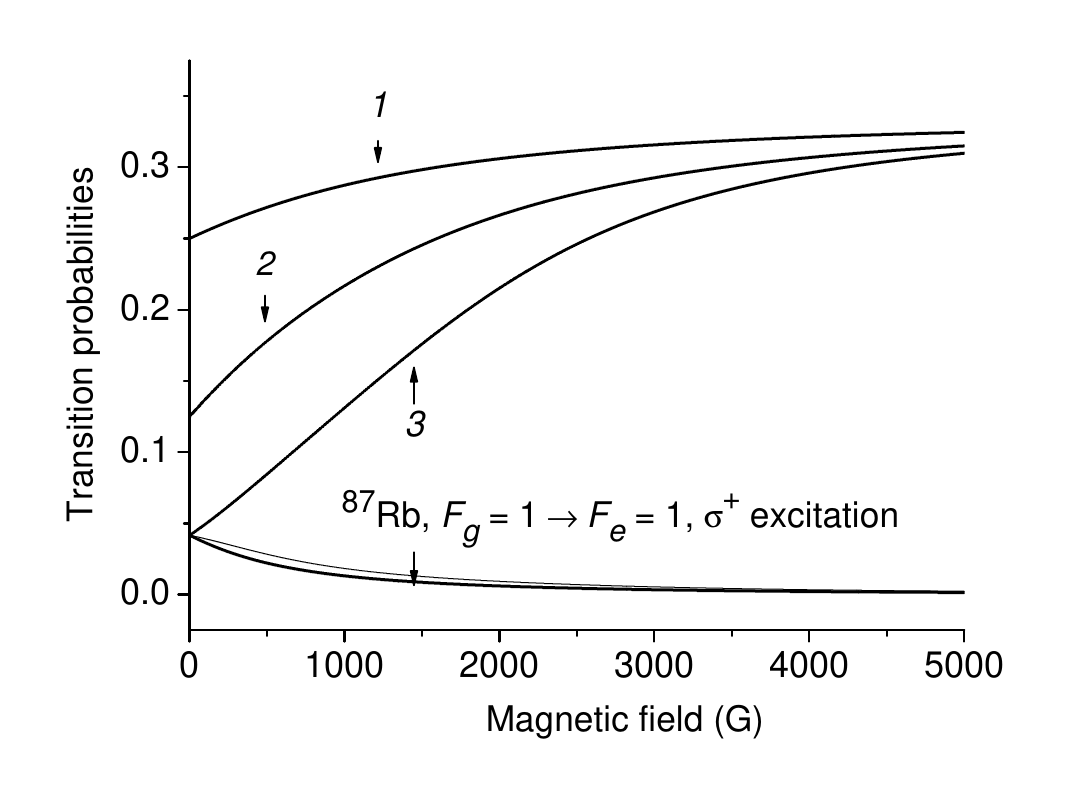}
}
\label{fig:8a}
} \hspace{0.5cm}
\subfigure[]{
\resizebox{0.4\textwidth}{!}{\includegraphics{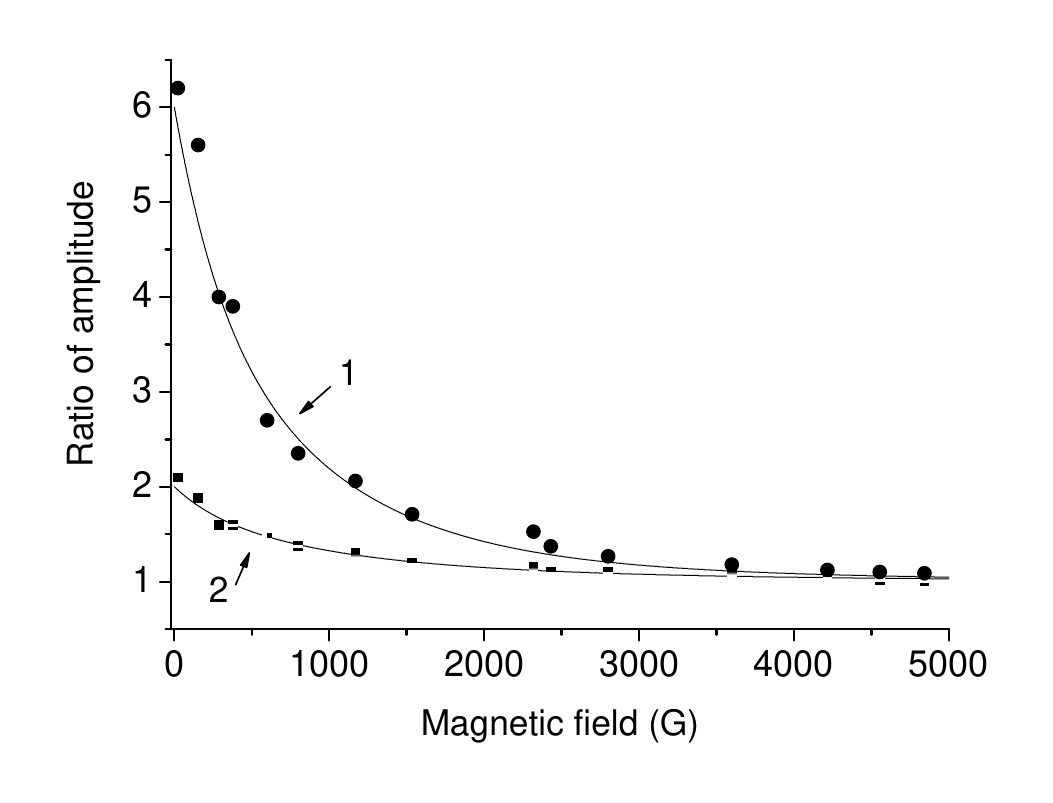}
}
\label{fig:8b}
}
\caption{(a) The probability for the atomic transition $1, 2, 3$ in the case of $\sigma^+$ 
excitation versus magnetic field (theory), (b) curve \textit{1} shows the ratio of the 
amplitudes $A(1)/A(3)$ and curve \textit{2} shows the ratio of the amplitudes $A(1)/A(2)$ versus $B$.}
\end{figure}
Fig. \ref{fig:8a} shows the probability for the atomic transitions \textit{1, 2, 3} (i.e. the 
atomic line intensity) for the case of $\sigma^+$ excitation versus magnetic field for (theory) 
versus $B$. However, in the experiment it is more convenient to measure the ratio of the VSOPs 
amplitudes $A1$, $A2$ and $A3$ of transitions \textit{1, 2} and \textit{3} versus $B$ (shown 
in Fig. \ref{fig:8b}), since the absolute value of the VSOP amplitude depends on laser intensity, 
NTC temperature, etc. Note that for $B\approx0$ the ratios $A1 : A2 : A3 = 6 : 3 : 1$ , while for 
large $B$, these ratios become $A1 \approx A2 \approx A3$. Also, as it is seen for $B$ up to 5000 G 
the VSOP amplitude $A1$ is increasing, which makes it convenient to use VSOP with label \textit{1} 
for a magnetic field measurement.
It is obvious that a NTC with an oven can also be fixed on a table with a micrometer step so that 
the displacements of this system allow one to map strongly inhomogeneous magnetic fields. For a 
more successful mapping, the dimensions of the NTC with oven could, in principle, be decreased 
further; i.e., a conducting and optically transparent deposited layer can replace the oven and 
the window thickness can be reduced to 100 $\mu$m with the reduction of the transverse dimensions 
of the NTC down a few millimeters.
It should be noted that, with regard to sensitivity, the magnetometer based on LZT is far below 
magnetometers based on coherent processes \cite{Budker_1,Budker_2}, but has advantages in the 
measurement of strong and gradient magnetic fields.
Note also that the spectral resolution achieved by LZT may also be realized with the use of 
well-collimated atomic beams (by means of 3 - 4 m vacuum pipes where the beam will be formed) 
or plants for cooling of atoms. However, employment of the atomic-beam technology or atom cooling 
is very complicated and expensive problem, whereas LZT requires only available cheap diode lasers 
and a nanocell filled with alkali metal.

\section{Conclusion}
The "$\lambda$-Zeeman technique" is shown to be a convenient and robust tool for the study of 
individual transitions between the Zeeman sublevels of hyperfine levels in an external magnetic 
field of 10 - 5000 G (taking into account previously obtained results of the LZT use for the 
range of 10 - 2500 G). LZT is based on NTC resonant transmission spectrum with thickness $L = \lambda$, 
where $\lambda$ is the resonant wavelength (794 nm) for $D_1$ line of the Rb. Narrow VSOP 
resonances (of $\sim 20 - 30$ MHz linewidth) in the transmission spectrum of the NTC are split 
into several components in a magnetic field; their frequency positions and transition 
probabilities depend on the $B$ field. Examination of the VSOP resonances formed in a nanometric-thin 
cell  allows one to obtain, identify, and investigate the atomic transitions between the Zeeman 
sublevels in the transmission spectrum of the ${}^{87}$Rb $D_1$ line in the range of magnetic 
fields 10 - 5000 G. Nanometric-thin column thicknesses ($\sim$ 794 nm) allow of the application 
of permanent magnets, which facilitates significantly the creation of strong magnetic fields. 
The results obtained show that a nano - magnetometer in the range of 10 - 5000 G with a local 
spatial resolution of $\sim$ 794 nm can be created based on a NTC and the atomic transition of 
the ${}^{87}$Rb $D_1$, $F_g = 1, m_F = +1 \rightarrow F_e = 2, m_F = +2$. This result is 
important for mapping strongly inhomogeneous magnetic fields.
LZT can be successfully implemented also for the case of Cs, Na, K, Li atoms. Experimental 
results are in a good agreement with the theoretical values.

\section{Acknowledgements}
The authors are grateful to A. Sarkisyan for his valuable participation in fabrication of the 
NTC as well as to A. Papoyan and A. Sargsyan for useful discussions. Research conducted in the 
scope of the International Associated Laboratory IRMAS. Armenian team thanks for a research 
grant Opt 2428 from the Armenian National Science and Education Fund (ANSEF) based in New York, USA.












\end{document}